# Exposure of the Hidden Anti-Ferromagnetism in Paramagnetic CdSe:Mn Nanocrystals

*Shou-Jyun Zou, Sheng-Tsung Wang, Ming-Fan Wu, Wen-Bin Jian, Shun-Jen Cheng\**

Department of Electrophysics, National Chiao Tung University, Hsinchu 300, Taiwan

\*Corresponding author e-mail: sjcheng@mail.nctu.edu.tw

We present theoretical and experimental investigations of the magnetism of paramagnetic semiconductor CdSe:Mn nanocrystals and propose an efficient approach to the exposure and analysis of the underlying anti-ferromagnetic interactions between magnetic ions therein. A key advance made here is the build-up of an analysis method with the exploitation of group theory technique that allows us to distinguish the anti-ferromagnetic interactions between aggregative $Mn^{2+}$ ions from the overall pronounced paramagnetism of magnetic ion doped semiconductor nanocrystals.  By using the method, we clearly reveal and identify the signatures of anti-ferromagnetism from the measured temperature dependent magnetisms, and furthermore determine the average number of $Mn^{2+}$ ions and the fraction of aggregative ones in the measured CdSe:Mn nanocrystals.

Magnetic ion doped semiconductor nanocrystals (NCs) have persistently drawn increasing attention over years owing to their application potential in spintronics and nano-magnetics.[1-4] In general, the magnetism of a semiconductor NC doped with iso-electronic magnetic ion dopants, *e.g.* CdSe:Mn NCs, is established mainly by the paramagnetism of magnetic ion dopants themselves, which are yet possibly subjected to additional anti-ferromagnetic (AFM) interactions if some of them are aggregated.[5-10]

Such an anti-ferromagnetism underlying in a paramagnetic semiconductor is however difficult to be fully exposed since it is significant only between nearest neighbour (NN) magnetic ions,[10-12] being a portion of randomly distributed magnetic ion dopants, and likely smeared out by the overall paramagnetism. Nevertheless, the AFM interactions between magnetic ions in magnetic semiconductors have been extensively thought to significantly affect the magnetic features, including the reduced effective Mn concentration,[13,14] fast spin relaxation,[15] and low Curie temperature in the ferromagnetic phase.[16-19]

As a common observed magnetic feature, the number of substitutional magnetic ions in a magnetic semiconductor NC estimated from the measured magnetization is usually lower than the real supplied amount of magnetic ion dopants. Such a discrepancy could be simply attributed to the loss of magnetic ion dopants in the synthesis processes but also possibly results from the underlying AFM. To reveal the latter origin, one must be able to identify the statistical distribution of the magnetic ions, or at least to distinguish the weakly interacting distant magnetic ions and the aggregative ones that are interacting anti-ferromagnetically. Yet, such a delicate consideration for magnetic ion dopants statistically distributed in a NC is actually difficult to achieve both experimentally and theoretically.

Practically, one often mixes the two kinds of magnetic ions (both of spatially apart and aggregative magnetic ions) in a magnetic semiconductor and then model the resulting magnetism phenomenologically, in terms of tunable effective magnetic ion concentration, on the base of mean field theory (MFT).[9,11,13,20] For magnetic ion doped semiconductor NCs, the validity of the MFT, where all individual magnetic ion spins are smeared out as a single continuous field, still remains as an open question. On the other hand, the advanced exact diagonalization (ED) technique allowing for considering magnetic ions individually and yielding reliable results is however applicable only for small number of magnetic ions ($N < 10^1$).[21-23] The lack of appropriate theoretical tools for analysis of magnetic NCs (especially the ones with the moderate number ($N \sim 10^1 - 10^2$) of Mn ions as studied in this work) sets an obstacle for more exploration into those intriguing magnetic nanostructures.

In this work, we present theoretical and experimental investigations of the magnetism of CdSe:Mn nanocrystals (NCs) with the number of $Mn^{2+}$ ions ranged from few to over seventy. To reveal the underlying AFM in the paramagnetic NCs, we propose an efficient approach to the exposure and analysis of the hidden AFM in the paramagnetism of CdSe:Mn nanocrystals. A key advance made in our theoretical treatment is the build-up of a solvable model which well preserves Mn spins individually and the exploitation of group theory technique for efficient counting of tremendously high degeneracies of the energy spectrum. With the theoretical aids, we clearly reveal and identify the signatures of AFM from the measured temperature dependent magnetization of CdSe:Mn NCs with the great number of Mn ions as many as ~75. Furthermore, we are able to estimate the average number of substitutional Mn ions and even infer the fraction of the aggregative Mn ions therein. By extending the employed model, the electrically charging effects on the enhancement of the magnetism of CdSe:Mn NCs are also studied and discussed.

**RESULTS AND DISCUSSION**

In this study, five samples (MCS504, MCS508, MCS515, MCS804, MCS815) of spherical $Cd_{1-x}Mn_xSe$ NCs of two different sizes (the diameters of $d = 5$ nm and $d = 8$ nm) with various Mn-concentrations ($x_{dop}$ =0.375%, 0.75% and 1.5%) were prepared using high-temperature organic solution approach.[24] Table 1 lists the averaged diameters ($d$), estimated Mn-concentration ($x_{dop}$), and the corresponding averaged number of Mn ions per NC ($N_{dop}$) for the all samples. The Mn-concentrations $x_{dop}$ are estimated according to the supplied dosage of the doping Mn dopants and controllable during the input for NC synthesis. Behind the specific values of $x_{dop}$, the Mn ions actually vary in number NC by NC and, based on the non-destructive magnetization measurements, no further information can account for their spatial distributions. Figure 1a depicts a CdSe:Mn NC in a NC ensemble doped with statistically distributed Mn ions, which as an example consists of four spatially apart Mn ions (referred to as distant Mn ions hereafter) and four spatially close ones (referred to as aggregative Mn ions). One can find the basic descriptions for the method of sample preparation and magnetization measurements in the section "Method" at the end of this article. Section S3 of Supporting Information (SI) provides more detailed information about sample synthesis and characterizations.

**Table 1. Data of the measured CdSe:Mn nanocrystal samples.**[a]

| Sample No. | $d$ | $N_{dop}$ | $x_{dop}$ | $N_{eff}$ | $x_{eff}$ | $N_T$ | $x_T$ |
|---|---|---|---|---|---|---|---|
| **MCS504** | 5nm | 4.5 | 0.375% | 3.0 | 0.25% | 3 | 0.25% |
| **MCS508** | 5nm | 9.0 | 0.75% | 5.4 | 0.45% | 6 | 0.50% |
| **MCS515** | 5nm | 18.3 | 1.5% | 6.7 | 0.55% | 8 | 0.66% |
| **MCS804** | 8nm | 18.5 | 0.375% | 4.2 | 0.09% | 8 | 0.16% |
| **MCS815** | 8nm | 75.0 | 1.5% | 45 | 0.90% | 75 | 1.50% |

[a] The data include the averaged diameter of NC ($d$), averaged numbers of doping $Mn^{2+}$ per NC ($N_{dop}$, $N_{eff}$, and $N_T$) estimated by the supplying dopant amount, the measured $B$-dependent and the measured $T$-dependent magnetisms, respectively (See text for details). The corresponding Mn

concentrations ($x_{dop}$, $x_{eff}$, and $x_T$) are deduced from the ratio of the estimated numbers of Mn ion and that of total Cd lattice sites ($N_{Cd}$) of NC, *i.e.* $x_{\text{dop/eff/T}} \equiv N_{\text{dop/eff/T}}/N_{\text{Cd}}$.

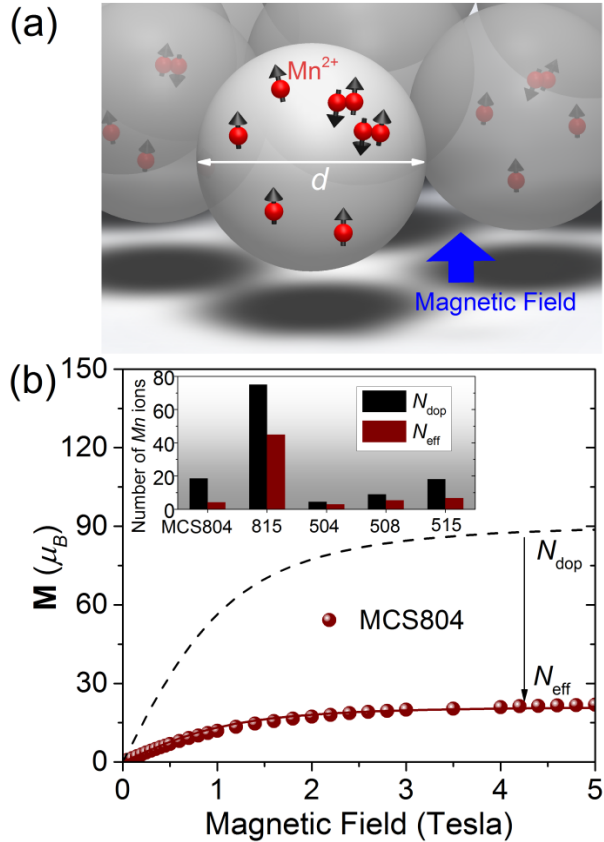

**Figure 1.** (a) Schematics of the measured CdSe:Mn nanocrystals (NC's) doped with $Mn^{2+}$ ions that are statistically positioned. (b) Measured magnetizations as functions of the magnetic field of the CdSe:Mn NC ensembles, MCS804. The solid lines (dashed lines) are the magnetizations of ideal paramagnets described by $\boldsymbol{M} = N_{\text{eff}}\, g_{Mn}\mu_B J B_J(\frac{g_{Mn}\mu_B B}{kT})$ with $N_{\text{eff}}$=4.2 (with $N_{\text{dop}}$=18.5). The inset shows the statistics data of the averaged numbers ($N_{\text{dop}}$'s) of Mn ions per NC that are estimated from the supplying dopant amounts in chemical synthesis processes and the numbers ($N_{\text{eff}}$'s) of magnetically effective Mn ions per NC that are deduced from the measured

magnetizations for the five NC samples. One notes that for all samples the $N_{\text{eff}}$'s are shown smaller than the $N_{\text{dop}}$'s.

**Experimental observations.** Figure 1b presents the measured averaged magnetization per NC of the sample MCS804 as a function of the magnetic field, which is shown well fitted by an ideal paramagnetic magnetization $\boldsymbol{M} = N_{\text{eff}} g_{Mn} \mu_B J B_J(\frac{g_{Mn}\mu_B B}{kT})$ but with the substitution of another effective number of Mn ions, $N_{\text{eff}}$=4.2, where $B_J(x)$ is the Brillouin function, $J=5/2$ is the spin of Mn ion, $g_{Mn} = 2.0$ is the g-factor of Mn ion, $B$ is the external magnetic field, $k$ is the Boltzmann constant, $T$ is the temperature, and $\mu_B$ is the Bohr magneton. Note that the corresponding effective Mn concentration, $x_{\text{eff}} \equiv \frac{N_{\text{eff}}}{N_{\text{Cd}}} \approx 0.09\%$, is significantly lower than that of supplied Mn ion dopants, $x_{\text{dop}} \equiv \frac{N_{\text{dop}}}{N_{\text{Cd}}} \approx 0.375\%$, where $N_{\text{Cd}}$ is the total number of Cd lattice sites in a CdSe:Mn NC.

To highlight the underestimated Mn concentration $x_{\text{eff}}$, figure 1b compares the observed magnetizations fitted with $x_{\text{eff}}$ and that one yielded by the Brillouin function formalism with the substitution of $x_{\text{dop}}$. The drastic quantitative difference between the two results indicates significant underestimation of the Mn concentration from the analysis of *B*-dependent magnetization, a common observed feature of magnetic semiconductor systems.[13,14] Table 1 compares the estimated Mn-concentrations ($x_{\text{dop}}$ and $x_{\text{eff}}$) by the two means for the all NC samples and confirms the underestimation of $x_{\text{eff}}$ and $N_{\text{eff}}$ as general phenomena. Graphically, the inset of figure 1b presents the two estimated average numbers of Mn ions per NC, $N_{\text{dop}}$ and $N_{\text{eff}}$, for all measured samples. The possible causes of the reduced $x_{\text{eff}}$ and $N_{\text{eff}}$ could be the loss of magnetic ion dopants during the synthesis processes or the existence of Mn-aggregations which are AFM interacting and likely magnetically inactive. In general, the both speculated causes are hardly distinguished and confirmed owning to the lacking of the detailed information about the number and distribution of the substitutional magnetic ions in a magnetic NC. Below we shall present a solvable model and an analysis method for revealing and identifying the signatures of AFM established from aggregated Mn ions from the measured temperature-dependent magnetism of Mn-doped semiconductor NCs.

**Model.** To conduct a physical analysis for the measured CdSe:Mn NCs with the moderate number of Mn ions (ranged from some to tens), we encounter the restrictions of the use of the ED and MFT methods that, as previously addressed in the introductory section, are applicable only for magnetically doped nanostructures with small number of magnetic ions or many magnetic ions that are homogeneously distributed, respectively. Here, we attempt to compromise the ED and MFT methods, and propose a model Hamiltonian that allows us to deal with, individually, statistically distributed Mn ions and is solvable for magnetic nano-structures with arbitrary number of Mn ions. The developed model makes the use of the fact that Mn-Mn interactions are essentially short-ranged and simplifies the position-dependent Mn-Mn interactions as fixed coupling constants by taking into account only the AFM interactions between *aggregative* Mn ions. Thus, the model Hamiltonian for a magnetic NC with $N$ Mn ions that are composed of $P$ distant and $Q$ aggregative Mn ions ($N=P+Q$) reads

$$H_M^{\text{eff}} = -g_{Mn}\mu_B B \sum_{i=1}^{N} m_i^z + \frac{J_M}{2} \sum_{i(\neq j)=1}^{Q} \sum_{j=1}^{Q} \vec{m}_i \cdot \vec{m}_j \qquad (1)$$

where $\vec{m}_i$ ($m_i^z$) is the spin ($z$-component of the spin) of the $i$-th Mn ion and $J_M$ is the constant of effective Mn-Mn coupling between the $Q$ aggregative Mn ions. The first term in Eq.(1) arises from the couplings between the external magnetic field and the *all N* paramagnetic ion dopants, while the second term represents the AFM interactions involving only the *Q aggregative* Mn ions. In the model, we assume that all Mn's in an aggregation are subjected to the equal effective Mn-Mn interactions so as to preserve the high permutation symmetry of the spin interactions that is needed later for efficient degeneracy counting using the group theory technique. However, under the assumption, the total strength of the effective anti-ferromagnetic Mn-Mn interactions could be overestimated. To fix the problem, in the use of the model one needs to appropriately rescale the strength of the effective Mn-Mn interactions according to the specific types of Mn aggregations existing in the NC. Statistically, as shown in Ref.[10], the most probable type of Mn ion aggregation could be paired Mn ions (*i.e.* dimer). Thus, throughout this work we shall consider the *Q* Mn ions paired as *Q/2* dimers, and estimate the effective AFM coupling constant by the mean value as $J_M \approx \frac{\frac{J_{MM}^{(0)} Q}{2}}{\frac{Q(Q-1)}{2}} = \frac{J_{MM}^{(0)}}{Q-1}$ meV, where $J_{MM}^{(0)} = 0.5$ meV is the coupling constant between nearest neighbour Mn ions in CdSe.[7,8,25] The model of Eq.(1) is extendible for more delicate consideration of various types of Mn aggregations (dimers, trimers, *etc.*) by including more corresponding spin-spin coupling terms and solved in similar manner. However, for the purpose to confirm and physically analyze the existing AFM behind the experimental observations, the model of Eq.(1) should be sufficiently useful.

Since the Hamiltonian of Eq.(1) commutes with the operators of total spin of all Mn ions $\vec{M}_N \equiv \sum_{i=1}^{N} \vec{m}_i$, $M_N^2$ and $M_N^z$, and those of total spin of all AFM-interacting Mn ions $\left(\vec{M}_Q \equiv \sum_{i=1}^{Q} \vec{m}_i\right)$, $M_Q^2$, the eigen-states can be represented by $|M_N, M_Q, M_N^z>$. Correspondingly, eigen-energies are solved as

$$E(M_N, M_Q, M_Z^z) = \frac{J_M}{2}\left(M_Q(M_Q+1) - \frac{35}{4}Q\right) - M_N^z g_{Mn} \mu_B B \quad (2)$$

According to Eq.(2), the magnetization and magnetic susceptibility of a Mn-doped NC supposedly can be calculated, which are $\boldsymbol{M} = -\left(\frac{\partial F}{\partial B}\right)_T = kT(\frac{\partial \ln Z}{\partial B})$ and $\chi = \frac{\partial \boldsymbol{M}}{\partial B}$ defined in terms of the partition function,

$$Z = \sum_{M_N, M_Q, M_N^Z} n_{(M_N, M_Q, M_N^Z)} e^{-\frac{E(M_N, M_Q, M_N^Z)}{kT}} \quad (3)$$

as a function of the level energies $E(M_N, M_Q, M_N^Z)$ and the degeneracies $n_{(M_n, M_Q, M_N^Z)}$ as well. In practice, the degeneracies $n_{(M_n, M_Q, M_N^Z)}$ are usually tremendously high (which are typically $n \gg 10^2$ for few Mn ions, and even as many as $n > 10^{36}$ for the number of Mn ions ~75, as shown in figure S1 of SI) so as to hardly count out either analytically or numerically. One should note that, even though the model Hamiltonian of Eq.(1) is solvable, the magnetism of a Mn-doped NC cannot be calculated if the degeneracies of energy spectrum remain unknown. The theoretical model and techniques presented here are in general applicable for any magnetic nanostructures containing finite number of magnetic ions, including other colloidal nanostructures and self-assembled quantum dots under extensive studies.[26-29]

A key theoretical progress further made here is the exploitation of group theory technique to make countable the degeneracies $n_{(M_n, M_Q, M_N^Z)}$ of NCs with *arbitrary* number of magnetic ion dopants, with no need of heavy numerical computation. The details of group theory technique used for counting the high spin degeneracies are presented in SI. By using the method, the complete energy spectra, including the level energies and degeneracies, of the measured magnetic NCs (including the ones doped with maximally 75 Mn ions per NC) in this work are calculated.

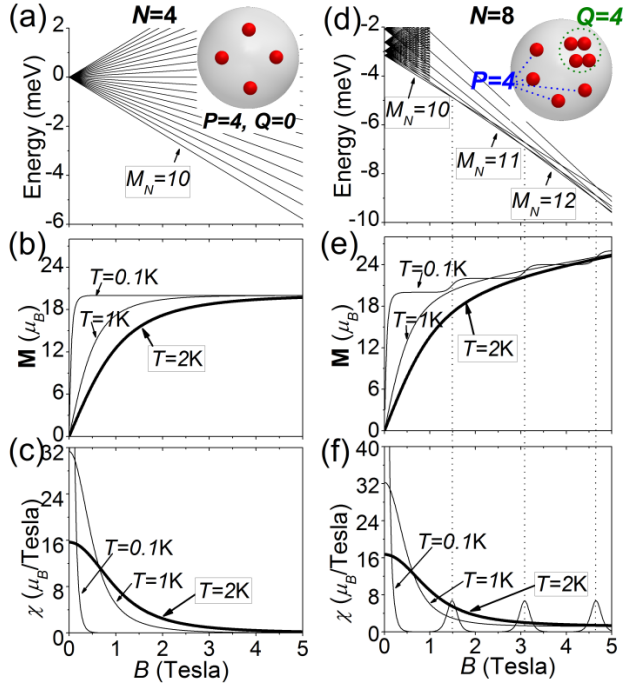

**Figure 2.** (a)-(c): Calculated energy spectrum, magnetization, and magnetic susceptibility of a magnetic CdSe:Mn NC with four distant Mn ions ($P=4$, $Q=0$). (d)-(f): Calculated results for the same magnetic CdSe:Mn NC but with four distant Mn ions ($P=4$) and four additional aggregative Mn ions ($Q=4$). Due to AFM interactions between the aggregative Mn ions, the ground states of the CdSe:Mn NC possess the minimized total angular momentum ($M_N=10$) at $B=0$ and then undergo successive increments of angular momentum ($M_N=10,11,12...$) with increasing the applied magnetic field.

**Analysis of *B*-dependent magnetism**. In theoretical analysis, we begin with the *B*-dependent magnetism of CdSe:Mn NCs doped with distant paramagnetic Mn ions only (*i.e.* $P\neq0$, $Q=0$), among which the Mn-Mn interactions are nearly vanishing. Figure 2a, 2b and 2c show the calculated energy spectrum, magnetization and magnetic susceptibility as functions of the magnetic field $B$ for the magnetic NC of diameter $d=5$ nm, doped with four distant Mn ions ($P$=4, $Q$=0, and $N$=4). In figure 2a, the ground states of the four-Mn doped NC are shown to be those of the fixed (maximum) angular momentum $M_N$=10 against the varied magnetic field. In the cases, the effective Hamiltonian Eq.(1) can be recovered to that of ideal paramagnet with $\boldsymbol{M} = g_{Mn}(NJ)B_J(\frac{g_{Mn}\mu_B B}{kT})$ with $J = \frac{5}{2}$ and $N$=4. In the low field limit $\left(\frac{g_{Mn}\mu_B B}{kT} \ll 1\right)$, $\boldsymbol{M} \approx J(J+1)N\frac{(g_{Mn}\mu_B)^2 B}{3kT}$ and $\chi \approx J(J+1)N\frac{(g_{Mn}\mu_B)^2}{3kT}$ well obeys the Curie's law.[30,31] Remarkably, the latter formalism shows that the product of magnetic susceptibility and temperature, $\chi T \propto J(J+1)N$, is invariant with varied temperature and reflects the magnitude of the total angular momentum $J$.

Next, let us introduce additional AFM-interacting aggregative Mn ions into the Mn-doped NCs (*i.e.* $Q \neq 0$). Figure 2d shows the energy spectrum of the same NC as considered in Figure 2a but with four additional AFM-interacting Mn ions ($P$=4, $Q$=4 and $N$=8). At zero or sufficiently small magnetic field, the total spin of the $Q$ aggregative Mn ions, that are subjected to AFM interactions, are vanishing and the total angular momentum of the ground states of the NC remains the same ($M_N$=10) as that of the NC with $P$=4 and $Q$=0 shown in figure 2a. Figure S1 shows the degeneracies (at the scales of $10^2$~$10^3$) of the low-lying states (ordered by energy and also labelled with their quantum numbers $M_N$) of the eight-Mn doped NC at zero magnetic field, which are calculated using the efficient group theory technique as presented in SI and confirmed numerically. In contrast to the NC with $Q$=0 (see figure 2a), the ground states of the magnetic NC with $Q$=4 AFM interacting Mn ions (see figure 2d) undergo successive increments of the angular momentum ($M_N$ =10, 11, 12, …) with increasing $B$, corresponding to the increments of the total spin of the aggregative Mn ions ($M_Q$ =0, 1, 2, …). As well studied previously by Refs.[10,12,32], the $B$-driven successive increments of the angular momentum of a AFM system result in the step feature of *M* and oscillating $\chi$ over a wide range of *B* at low *T*, as also seen in figure 2e and 2f.

However, such an essential difference between the magnetic features of the NCs with and without AFM interacting Mn ions at low *T* is likely smeared out by thermal fluctuations. As shown in figure 2e and 2f, the step feature of *M* and the corresponding oscillating $\chi$ of the NCs doped with aggregative Mn ions are completely faded out as the temperature is raised to *T*=2 K, and turn out to be indistinguishable from those of the NC without any aggregative Mn ions (see curves for *T*=2 K in figure 2b and 2c).

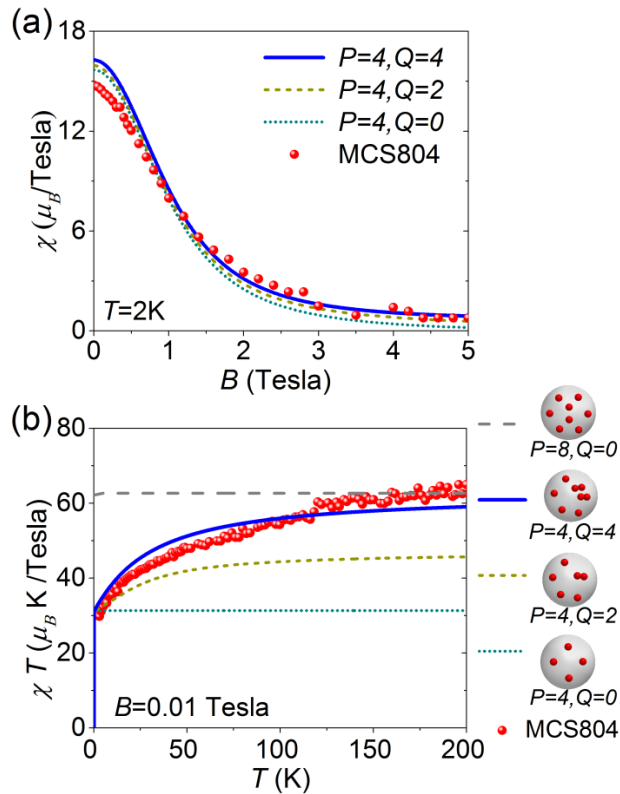

**Figure 3.** (a) Measured susceptibility $\chi$ of the sample MCS804 at *T*=2 K as a function of magnetic field and the simulated results of the NCs with fixed four distant Mn ions (*P*=4) but various numbers of aggregative Mn ions (*Q*=0, 2, 4). The simulation shows that, even at the low

temperature, the *B*-dependent magnetic susceptibility of a NC is insensitive to the variation of the number of AFM interacting aggregative Mn ions, and hard to evidence the underlying AFM in the paramagnetic NCs with Mn-aggregations. (b) Measured product of magnetic susceptibility and temperature, $\chi T$, of the sample MCS804 at *B*=0.01Tesla versus the varied temperature. The simulated results of the NCs with *P*=4 and *Q*=0,2,4, are presented to fit the experimental data. One notes that the *T*-dependence of the measured $\chi T$ are very sensitive to the portions of the distant and aggregative Mn ions, and by fitting the *T*-dependent $\chi T$ one can infer the most possible (*P*,*Q*) of a magnetic ion doped NC. In this case, the simulated result for *P*=4 and *Q*=4 show the best fitting.

Figure 3a shows the measured magnetic susceptibility of the MCS804 NC-sample at *T*=2 K as functions of the magnetic field, and, for comparison, the theoretical results of the magnetic NCs with fixed *P*=4 distant Mn ions and varied number (*Q*=0,2,4) of aggregative Mn ions. One sees that all of the NCs undoped or doped with different number of aggregative Mn ions exhibit similar *B*-dependences of the magnetisms at *T*=2 K. Therefore, simply from the *B*-dependent magnetism, one cannot judge the existence and even infer the number of aggregative Mn ions in a magnetic ion doped NC.

**Analysis of *T*-dependent magnetism**. Alternatively, we suggest to study the AFM established between aggregative Mn ions in the magnetic NCs much better by analyzing the product of $\chi T$ as a function of the temperature *T*. Another way often advised by standard AFM theory is by means of examining the *T*-dependence of the magnetic susceptibility, $\chi$ vs. *T*. Nevertheless, as presented in figure S2 and discussed in SI the usefulness of the $\chi - T$ analysis is limited for the CdSe:Mn NCs studied here since only part of Mn ions in Mn-doped NCs are aggregated and AFM-interacting.

Figure 3b presents the measured $\chi T$ 's of the MCS804 sample at the varied temperatures from $T=2$ K to 200K, which are not shown invariant with respect to the varied $T$ as predicted by the Curie's law. To understand the observed feature, we calculate the $\chi T$'s, as functions of the $T$, for the NCs with various mixtures of distant and aggregative Mn ions, ($P=4$, $Q=0$), ($P=4$, $Q=2$), ($P=4$, $Q=4$) and ($P=8$, $Q=0$). For the NCs without any aggregative Mn ions (($P=4$, $Q=0$), ($P=8$, $Q=0$)), the values of the calculated $\chi T$ 's do remain unchanged with respect to the varied $T$, and are determined by the number of $P$. Adding aggregative Mn ions to the NC ($Q=2,4$), the calculated $\chi T$ 's turn out to be $T$-dependent, which are increasing with $T$ and gradually approaching the maximum values as $T>100$ K.

As mentioned previously, the magnitude of $\chi T$ of a magnetic system is associated with the total angular momentum of the system in the Curie's law. At extremely low $T$, because of the existing AFM interactions the angular momentum of the ground states of a magnetic NC with $Q \neq 0$ are minimized, where the spins of the individual aggregative Mn ions are anti-parallel. This leads to the lowest value of $\chi T$ at $T \rightarrow 0$, as shown in figure 3b for the cases of ($P=4$, $Q=2$), ($P=4$, $Q=4$). With increasing $T$, the NCs could thermally access more excited states of higher angular momentum, and the $\chi T$ gradually increases. At very high $T$ (such as $T >100$ K), all states of various angular momenta are nearly equally accessed and the $\chi T$ eventually approaches a saturated value. One thus realizes that, unlike the $B$-dependent magnetism, the $T$-dependences of the $\chi T$ 's of the NCs with different $P$ and $Q$ are quite distinct. This allows us to identify and distinguish AFM-interacting Mn-aggregations from the overall paramagnetism of a CdSe:Mn NC sample by examining the $T$-dependence of the measured $\chi T$.

In figure 3b, the theoretical curve of ($P=4$, $Q=4$) is shown to best fit the experimental data of the MCS804 sample. Correspondingly, the estimated Mn concentration from the fitting of the $T$- dependence of the measured $\chi T$ is $x_T = N_T / N_{Cd} = 0.16\%$, which lies between those, $x_{dop} = N_{dop} / N_{Cd} = 0.375\%$ and $x_{eff} = N_{eff} / N_{Cd} = 0.09\%$, that are estimated from the supplied amount of dopant and fitting of *B*-dependent magnetism, respectively. Following the same fitting analysis, the sample MCS515 [MCS815] is inferred to contain ~8 [~75] Mn ions composed of $P=5$ and $Q=3$ ($P=25$ and $Q=50$), yielding the corresponding Mn concentration $x_T=0.66\%$ [$x_T=1.50\%$], as shown in figure 4a and 4b [figure 4c and 4d]. Remarkably, the high degeneracies, as many as ~$10^{36}$ (see figure S1b), of the energy spectrum of the NCs doped with 75 Mn ions need to be known precisely in the calculation of magnetization and can be counted out only by using the efficient group theory technique. One also notes that the measured $\chi T$'s of the NC ensembles persistently descend with increasing the temperature $T > 100K$. The observed linearly descending feature of $\chi T$ could be attributed to the inherent diamagnetism of host material CdSe,[33-34] which was not considered in the model. The temperature-insensitive diamagnetism makes a negative constant offset to the total magnetic susceptibility and leads to a linear descending $\chi T$'s.[35]

Figure 5 summarizes and compares the estimated numbers of Mn ions per NC by different means. Among the three estimated numbers of Mn ions for a sample, $N_{dop}$'s always show the highest number since $N_{dop}$'s count not only the Mn ions aggregated but also the ones not really incorporated into the NCs. By contrast, the values of $N_{eff}$'s are always the smallest since $N_{eff}$'s count only the spatially distant Mn ions but might overlook all aggregative Mn ions that which are often magnetically inactive. As best estimated numbers, $N_T$'s lie usually between the numbers $N_{dop}$ and $N_{eff}$ for the same NC sample since $N_T$'s rule out the Mn ions not really doped into the NCs but take into account both of the spatially distant and aggregated Mn ions.

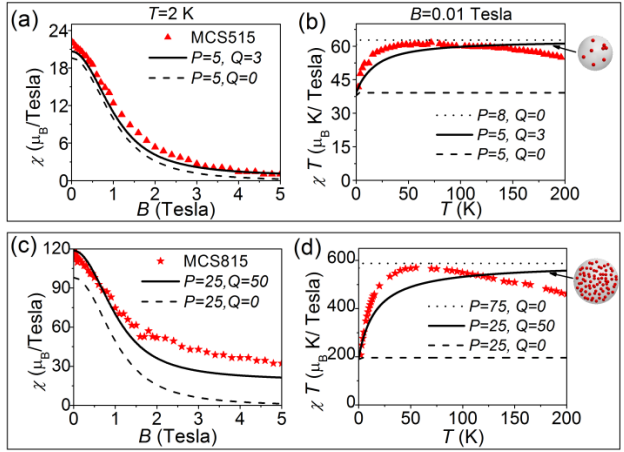

**Figure 4.** (a) Measured magnetic susceptibility $\chi$, as a function of $B$, of the samples MCS515 at $T$=2 K and (b) the measured $\chi T$, as a function of $T$, of the same samples at $B$=0.01 Tesla. Correspondingly, the theoretical results for the NCs with various numbers of distant and aggregative Mn ions are presented in the plots for fitting the experimental data. (c) and (d) show the measured and theoretical results of the MCS815 sample, composed of the NCs doped with high number of Mn ions, under the same conditions as (a) and (b). As results of fitting, the NCs in the MCS515 (MCS815) sample are inferred to contain, in average, 5 (25) distant Mn ions and 3 (50) aggregative Mn ions.

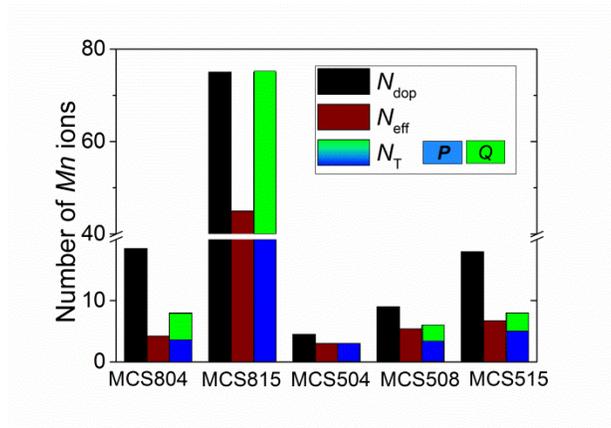

**Figure 5.** Statistics of the average numbers ($N_{dop}$, $N_{eff}$, and $N_T$) of Mn ions per NC of the CdSe:Mn NC samples MCS504, MCS508, MCS515, MCS804 and MCS815, which are estimated by three different means, respectively. The number of total Mn ions ($N_T$), composed of the aggregative Mn ions ($Q$) and distant Mn ions ($P$), for a NC is deduced from the best fittings of the $T$-dependence of the $\chi T$, as presented in Figures 3b, 4b and 4d.

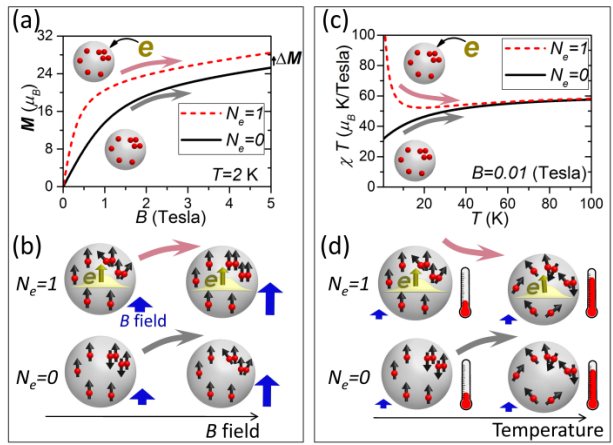

**Figure 6.** (a) Calculated magnetizations as functions of magnetic field $B$ of a magnetic NC of diameter $d=5$nm doped with eight Mn ions composed of ($P=4$, $Q=4$) uncharged (black solid line) and charged with a single electron (red dashed line). The inset (b) illustrates how the spins of aggregated Mn ions and the resulting magnetization of a magnetic NC increase with increasing $B$ no matter if whether the NC is charged or not. (c) Calculated product of temperature and magnetic susceptibilities, $\chi T$, of the un-charged (black solid line) and singly charged (red dashed line) magnetic NCs with fixed magnetic field $B=0.01$Tesla but varied temperature $T$. The inset (d) illustrates that the spin alignments of Mn ions and the resulting magnetism of un-charged and charged magnetic NC evolve in distinctive ways against the varied $T$.

**Charging effects.** Another advantageous feature of magnetic NCs is the charge-controlled magnetism. Charging a magnetic ion doped NC creates the quantum-confinement enhanced spin interactions between confined charged carriers and Mn ions which tend to maximize the magnetism of the NC. Experimentally, S. T. Ochsenbein *et al.* [36,37] have presented the charge-controlled magnetism of $Zn_{1-x}Mn_xO$ NCs. Extending the employed model for uncharged Mn-doped NCs, the Hamiltonian of a singly charged NC doped with many Mn ions is given by

$$H_{eM}^{eff} = -J_c \vec{s} \cdot \sum_{i=1}^{N} \vec{m}_i + \frac{J_M}{2} \sum_{i \neq j; i,j=1}^{Q} \vec{m}_i \cdot \vec{m}_j - I_z g_I \mu_B B \quad (4)$$

where the first term represents the magnetic couplings between the quantum confined electron and the all magnetic ions in the NC, the second one is the magnetic couplings between aggregative Mn ions, and the last one is the spin Zeeman term, $\vec{s}\ (s = \frac{1}{2})$ denotes the spin of charging electron, $\vec{I} = \vec{s} + \vec{M}_N$ ($I_z$) is the total spin (the *z*-component of the total spin) of electron and Mn ions, $g_I$ is the Lande *g*-factor.

The strength of the magnetic coupling is quantum-confinement engineerable and estimated by $J_c = 6 J_{eM}^{(0)} / (\pi d^3)$ meV with $J_{eM}^{(0)} = 10.8$ meV nm$^3$ in the spherical hard wall model for CdSe:Mn NCs.[5,38] Since the effective Hamiltonian of Eq.(4) commutes with the spin operators, $I^2, I_z, M_N^2$ and $M_Q^2$, the eigen-states can be represented by the corresponding quantum numbers, *i.e.* $|I, M_N, M_Q, I_z>$, and the energy spectrum is explicitly solved as $E(I, M_N, M_Q, I_z) = -\frac{J_c}{2}\left(\frac{1}{2} \pm (M_N - \frac{1}{2})\right) + \frac{J_M}{2}\left(M_Q(M_Q + 1) - \frac{35}{4}Q\right) - I_z g_J \mu_B B$, where the up/down signs ($\pm$) correspond to $I = M_N + s$, and $I = M_N - s$ respectively. The magnetizations and magnetic susceptibilities of singly charged Mn-doped NCs can be calculated following the same approach used previously for un-charged CdSe:Mn NCs.

Figure 6a shows the calculated magnetizations of charged and uncharged CdSe:Mn NCs with $P=4$, $Q=4$ Mn ions as functions of the magnetic field. One sees that charging an electron onto a CdSe:Mn NC does make an overall increase of the magnetization, but remains a very similar $B$-dependence to that of un-charged one. This is because the electron-Mn couplings act as an additional effective field and polarize the spins of aggregative Mn ions. However, the electron-Mn couplings themselves are weakly $B$-dependent and do not alter much the $B$-dependent feature of the magnetization. Figure 6b illustrates why the magnetization of charged NCs is overall larger than that of uncharged NCs. Thus, *via* single measurements on the $B$-dependent magnetism of a magnetic NC, one still cannot judge if the NC is charged or un-charged and further identify the charging effects on the magnetism.

Again, we suggest that analyzing the $T$-dependence of the quantity $\chi T$ is an excellent way to identify the charging effect on the enhanced magnetism of a magnetic ion doped NC. Figure 6c shows the calculated $\chi T$'s of the uncharged and singly charged magnetic NCs with $P=4$ and $Q=4$ Mn ions. It is obviously observed that the charged CdSe:Mn NC exhibits an essentially different $T$-dependence of $\chi T$ from that of the uncharged NC. While the calculated $\chi T$ of an uncharged NC shows a monotonically increasing feature with increasing $T$, the same NC but charged with a single electron shows a much greater value of $\chi T$ at the lowest $T$ and then, contrarily, a dramatic decrease with increasing $T$. The contrast between the $T$-dependent magnetic features of $\chi T$ of a charged and an un-charged magnetic ion doped NCs allows us to unambiguously identify the charge-enhanced magnetism of a magnetic ion NC. Figure 6d illustrates how charging a magnetic NC leads to the drastically different $T$-dependence of $\chi T$. At low $B$ and $T$ (where $\mu_B B, kT < J_M \sim 0.5 \text{meV}$), the spins of the distant Mn ions are fully polarized by the external $B$-field while the total spin of the aggregative Mn ions is minimized due to the AFM interactions. In the charged NCs, the electron-Mn interactions that act as a huge effective field to Mn ions might turn the spins of the aggregative Mn ions partially polarized, as schematically shown by the NC for low $T$ and $N_e=1$ in figure 6d. As a result, the $\chi T$ of the charged NC at low $T$ is much higher than that of uncharged NC, as seen in figure 6c. With increasing temperature, the thermal effect tends to depolarize the polarized spins of the distant Mn ions and the aggregative ones as well in the charged NC, and the resulting $\chi T$ of decreases. Differently, the spins of the aggregative Mn ions in the uncharged NC are un-polarized due to the AFM interactions. Increasing temperature yet tends to polarize the spins of the aggregative Mn ions in the un-charged NC because the high spin states of the aggregative Mn ions lie at higher energy and are accessed thermally. Thus, even though the spins of distant Mn ions are subjected to the same thermal depolarization effect, the overall $\chi T$ slightly increases with increasing $T$, showing a totally different feature of $T$-dependence of $\chi T$ from that of the charged NC.

**CONCLUSION**

In summary, we present the theoretical and experimental investigations of the magnetism of CdSe:Mn nanocrystals. For physical analysis of the magnetic NCs doped with magnetic ion dopants whose spatial distribution and numbers are essentially random and statistical, we propose a solvable model Hamiltonian that can deal with the interacting Mn spins individually for arbitrary number of Mn ions and establish an analysis method for revealing the underlying AFM established between aggregated Mn ions in paramagnetic Mn-doped NCs. By fitting the observed temperature dependent magnetic susceptibilities $\chi T$ with theoretical calculations, we clearly reveal the signatures of anti-ferromagnetism established between short-range interacting Mn aggregation of CdSe:Mn nanocrystals that are usually well hidden in the pronounced paramagnetism and hard to be exposed. It is shown that in our magnetic NC samples about 0~60% of magnetic ions aggregate and significantly reduce the total magnetizations. Moreover, we also show that analyzing the *T*-dependence of the quantity $\chi T$ is an excellent way to identify the charge-controlled magnetism of a magnetic ion doped nanocrystal, being another remarkably advantageous feature of magnetic nanostructures in applications of spintronics and nano-magnetism.

**METHODS**

**Sample preparation.** The uniformity of NC size was controlled by growth time and further improved by a size selection procedure.[24] The crystalline structures, the morphology of the spherical shapes and the sizes of the NCs were confirmed from the high-resolution transmission electron microscope (TEM) images, as shown in the graphic in the graphical abstract. Accordingly, the NC-size distributions are confirmed to have a variation within a small standard deviation of only 7% for all the samples and the high-quality crystalline structures of NCs are not affected by the doping of Mn ions. It is also confirmed that the sizes of the NCs and the Mn-concentrations are uncorrelated. More detailed information about the sample synthesis and characterizations can be found in Section S3 in Supporting Information.

**Measurements.** Magnetizations of $Cd_{1-x}Mn_xSe$ NCs were measured by a superconducting quantum interference device (SQUID), over the ranges of temperature from 2 to 200K and of magnetic field from 0 to 5 Tesla. The as-grown $Cd_{1-x}Mn_xSe$ NCs stabilized with capping agents of both TOP (trioctylphosphine) and oleic acid, whose magnetizations are much smaller than bulk $Cd_{1-x}Mn_xSe$ and are negligible. To extract the magnetization of CdSe:Mn NCs from the measured data, we subtracted only the sample holder background (mainly the capsule), which is about $-1 \times 10^{-6}$ emu at 0.1 Tesla.

ACKNOWLEDGMENT

This work was supported by the National Science Council of Taiwan under Contract No. NSC-100-2112-M-009-013-MY2 and the Center for Interdisciplinary Science (CIS) at National Chiao Tung University. The authors thank Jiye Fang (State University of New York at Binghamton) for providing the nanocrystal materials. SJC is grateful to National Center for Theoretical Science for support.

ABBREVIATIONS

NC, AFM, NN, MFT, ED, TEM, SQUID, SI.

# Supporting Information

**SECTION S1**: Group theory technique for efficient counting of spin degeneracies.

This session presents the theoretical approach, based on the group theory,[1] to an efficient counting of the degeneracies of the spin eigen states $|M_N, M_Q, M_N^z>$ of magnetic ion doped NCs described by Eq.(1) in the main text, which are usually tremendously large even for a not great number of magnetic ions. The high degeneracies of the states $|M_N, M_Q, M_N^z>$ result from twofold spin permutation symmetries in Eq.(1), which lie in the first term for the $N$ Mn ions and the second one for the AFM-interacting $Q$-Mn aggregations in Eq.(1), respectively.

In the presence of a magnetic field, the $z$-component of the spin $M_N^z$ of the degenerate states $|M_N, M_Q, M_N^z>$ are resolved by the spin Zeeman energies and the degenerate states are split into $2M_N^z + 1$ sub-levels. As a result, only $M_N$ and $M_Q$ are relevant to the degeneracies $n_{M_N, M_Q, M_N^z}$ of each sub-levels for any specified $M_N^z$, which are needed in Eq.(S1) in the main text for the calculation of the magnetization.

For the magnetic NC with the $N$ spin-5/2 Mn ions, composed of $Q$ AFM-interacting Mn ions and ($N$-$Q$) non-interacting ones, the $6^N$-dimensional Hilbert space span by the spined eigen states can be decomposed as $\left(R_{\frac{5}{2}}\right)^Q \otimes \left(R_{\frac{5}{2}}\right)^{N-Q}$, where $R_k$ denotes an irreducible representation of spin $k$ described by SU(2) group.

Further, by using the tensor product decomposition formulae,[1] one can successively decompose the $6^N$-dimensional Hilbert space and formulate it as a linear combination of numerous subspaces $R_{M_Q M_N}$ for distinct $(M_N, M_Q)$,

$$\left(R_{\frac{5}{2}}\right)^Q \otimes \left(R_{\frac{5}{2}}\right)^{N-Q} = \sum_{M_Q} a_{M_Q} R_{M_Q} \otimes \left(R_{\frac{5}{2}}\right)^{N-Q} = \sum_{M_Q, M_N} a_{M_Q} a_{M_Q M_N} R_{M_Q M_N}$$

where $a_{M_Q}$ is the total number of the configurations with the same $M_Q$ of the $Q$-Mn aggregation and $a_{M_Q M_N}$ is the total number of the configurations with the same $M_N$ for a specified $M_Q$.

The degeneracies of a state $|M_N, M_Q, M_N^z>$ is therefore simply

$$n_{M_N, M_Q, M_N^z} = a_{M_Q} a_{M_Q M_N}$$

where $a_{M_Q}$ and $a_{M_Q M_N}$ can be straightforwardly obtained from the orthogonality relation for characters of SU(2),

$$a_{M_Q} = \frac{1}{\pi} \int_0^\pi \left(\lambda_{\frac{5}{2}}\right)^Q \lambda_{M_Q} (1 - cos\phi) d\phi$$

$$a_{M_Q M_N} = \frac{1}{\pi} \int_0^\pi \left[\left(\lambda_{\frac{5}{2}}\right)^{N-Q} \lambda_{M_Q}\right] \lambda_{M_N} (1 - cos\phi) d\phi$$

with $\lambda_k = \frac{sin[(k+\frac{1}{2})\phi]}{sin\frac{1}{2}\phi}$ being the character of spin-$k$ representation of SU(2).[1]

Figure S1 shows the calculated degeneracies of the low-lying states (ordered by energy and also labelled with their quantum numbers $M_N$) of the 8-Mn doped NC with $P=4, Q=4$ and 75-Mn doped NC with $P=25, Q=50$ at zero magnetic field. Note that the degeneracies of the NC with 75 Mn ions are as many as $\sim 10^{36}$, and countable only by using the efficient group theory technique presented in this section.

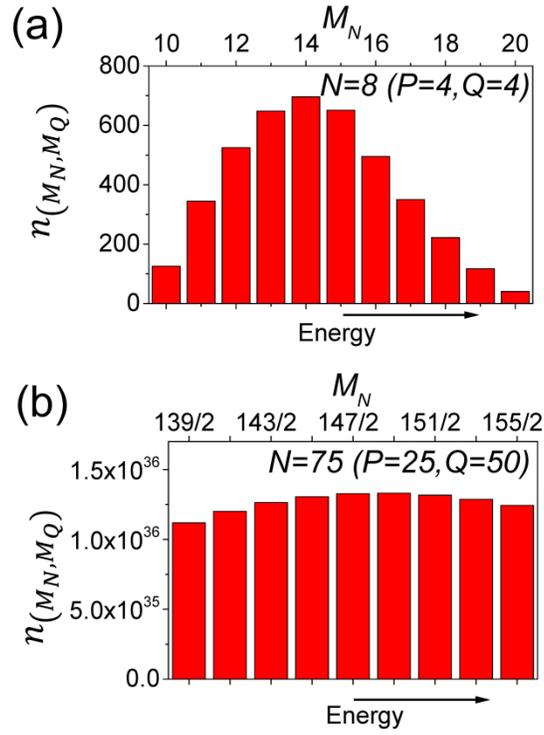

**Figure S1.** Statistics of the degeneracy numbers of the low-lying states of (a) a few-Mn doped CdSe:Mn NC with *P=4, Q=4*, and (b) a many-M doped CdSe:Mn NC with *P=25, Q=50*. Note that in the latter case of Mn-rich NC the degeneracies are as high as $\sim 10^{36}$.

**SECTION S2**: Analysis of temperature-dependent magnetic susceptibility.

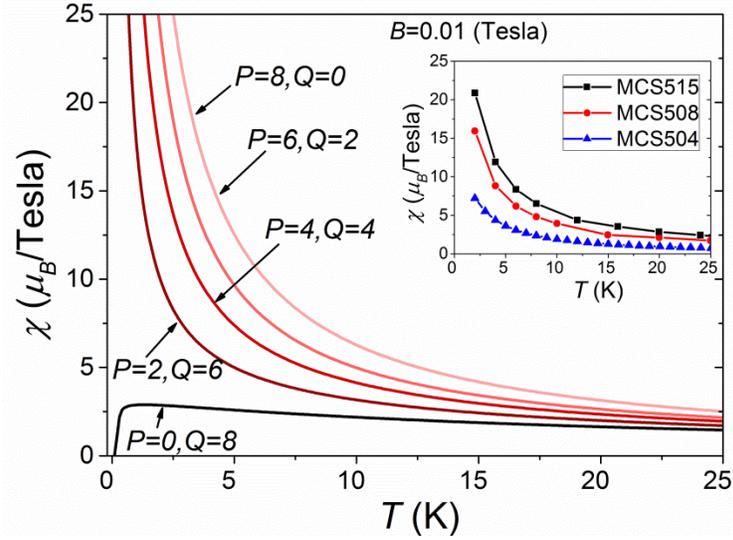

**Figure S2.** Calculated magnetic susceptibilities $\chi$ as functions of the temperature $T$ of CdSe:Mn NCs, at the low magnetic field $B=0.01$ Tesla, doped with eight Mn ions but in five different configurations, i.e. ($P=8$, $Q=0$), ($P=6$, $Q=2$), ($P=4$, $Q=4$), ($P=2$, $Q=6$), and ($P=0$, $Q=8$), respectively. It is clearly seen that only the NC full of AFM-interacting Mn ions ($P=0$, $Q=8$) shows a AFM-like T-dependence of magnetic susceptibility where a cusp feature is observed at $T\sim 1K$. Inset: Measured low-field $\chi$'s as functions of $T$ of the CdSe:Mn NC samples MCS504, MCS508, MCS515. Note that none of the measured $\chi$'s as functions of $T$ exhibit the AFM cusp feature. Nevertheless, one cannot exclude the existence of AFM-interacting Mn ions in the NCs according to the simulated results for $P \neq 0$ and $Q \neq 0$.

A standard way to identify an anti-ferromagnetism, as commonly advised by most textbooks on magnetism,[2,3] is by means of analysing the $T$-dependence of the magnetic susceptibility at the low magnetic field, which typically show a maximal cusp at a finite temperature (referred to as the Neel temperature) for an anti-ferromagnet. Nevertheless, we found that such a cusp feature of $\chi(T)$ does not emerges for the magnetic NCs studied in this work that are *partially* doped with AFM-interacting Mn ions. The measured low-field $\chi(T)$ as functions of temperature of the CdSe:Mn NC samples MCS504,MCS508,MCS515 are presented in the inset of figure S2. It is clearly shown that none of the measured $\chi(T)$ shows such AFM cusp features. Instead, the measured $\chi(T)$ are shown decreasing monotonically with increasing $T$, and behave as paramagnets.

To understand the experimental observations, we calculate the magnetic susceptibilities $\chi$ at $B=0.01$ Tesla, as functions of temperature $T$, of CdSe:Mn NCs doped with totally eight Mn ions but in different distributions, characterized by different numbers of distant and aggregative Mn ions, ($P=8$, $Q=0$), ($P=6$, $Q=2$), ($P=4$, $Q=4$), ($P=2$, $Q=6$), and ($P=0$, $Q=8$), as shown in figure S2. One notes that the typical AFM cusp feature of $\chi(T)$ emerges at a specific T only as the NC is full of aggregative magnetic ions, i.e. ($P=0$, $Q=8$). Once upon a NC contains distant paramagnetic Mn ions, that cusp feature disappears and instead a typical paramagnetic feature, i.e. a monotonically decaying $\chi$ with increasing $T$, is recovered. Therefore, as experimentally observed, one cannot confirm if there exist Mn aggregations or not by examining the the $T$-dependence of the magnetic susceptibility.

**SECTION S3**: Experimental details.

This section provides more information about the method of synthesis and characterizations of the CdSe:Mn nanocrystal ensembles under this study.

*Synthesis.* The measured CdSe:Mn nanocrystals were synthesized by high-temperature organic solution approach.[4,5,6] As described in Ref. 6, 2 mg of $Mn_2(\mu\text{-SeMe})_2(CO)_8$ and 0.7 ml of 1 M Se-trioctylphosphine solution were premixed with 3 ml of trioctylphosphine in a glove box. The premixed solution was rapidly injected into 20 g of trioctylphosphine oxide (90%) with 0.5 ml of 1 M cadmium(II) acetate in oleic acid at 310 °C under a flow of argon on an Schlenk line. The hot mixture was vigorously agitated at 260 °C for 1 min–1 h to produce different average-sized crystals. The growth of nanocrystals was then terminated by cooling it to room temperature. The same synthesis method was also employed in the preparation of PbSe:Mn nanocrystals in our previous studies.[7] The as-synthesized nanocrystals were subsequently refined by using a size-selection method.

*Characterizations.* The sizes, morphologies, and crystalline structures of the measured CdSe:Mn nanocrystals were estimated and observed by high-resolution transmission electron microscope (TEM) (See Figure S3).[6] The TEM measurements confirms the high crystalline quality of the nanocrystal samples. Besides, electron paramagnetic resonance (EPR) spectra of the nanocrystals were previously measured as shown in Figure S4 and Ref.6. As reported by Ref.6, the EPR measurement confirms the successful incorporation of $Mn^{2+}$ ions into the CdSe host material of nanocrystals and enables us estimate the averaged Mn-Mn interactions in the Mn-doped nanocrystals, showing sensitive dependences on the averaged Mn-Mn distance.

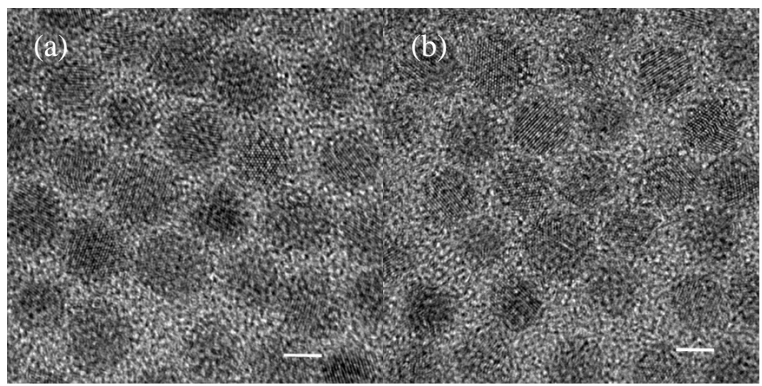

Figure S3. (a) TEM image of MCS504 with a scale bar of 2.5 nm. (b) TEM image of MCS804 with a scale bar of 5 nm.

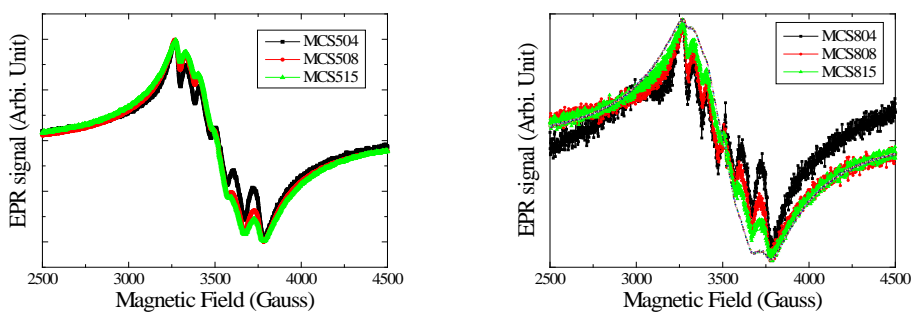

Figure S4. (a) EPR data of the CdSe:Mn nanorystals with the diameter of 5 nm. (b) EPR data of the CdSe:Mn nanocrystals with the diameter of 8 nm.